# Classification of attention performance post-longitudinal tDCS via functional connectivity and machine learning methods

Akash K Rao
*Applied Cognitive Science Laboratory*
*Indian Institute of Technology Mandi*
Mandi, India
0000-0003-4025-1042

Vishnu K Menon
*Applied Cognitive Science Laboratory*
*Indian Institute of Technology Mandi*
Mandi, India
0009-0007-9449-0934

Arnav Bhavsar
*School of Computing and Electrical Engineering*
*Indian Institute of Technology Mandi*
Mandi, India
arnav@iitmandi.ac.in

Shubhajit Roy Chowdhury
*School of Computing and Electrical Engineering*
*Indian Institute of Technology Mandi*
Mandi, India
src@iitmandi.ac.in

Ramsingh Negi
*Cognitive Control and Machine learning group*
*Institute of Nuclear Medicine and Allied Sciences*
Delhi, India
ramsingh@inmas.gov.in

Varun Dutt
*Applied Cognitive Science Laboratory*
*Indian Institute of Technology Mandi*
Mandi, India
varun@iitmandi.ac.in

*Abstract*—Attention is the brain's mechanism for selectively processing specific stimuli while filtering out irrelevant information. Characterizing changes in attention following long-term interventions (such as transcranial direct current stimulation (tDCS)) has seldom been emphasized in the literature. To classify attention performance post-tDCS, this study uses functional connectivity and machine learning algorithms. Fifty individuals were split into experimental and control conditions. On Day 1, EEG data was obtained as subjects executed an attention task. From Day 2 through Day 8, the experimental group was administered 1mA tDCS, while the control group received sham tDCS. On Day 10, subjects repeated the task mentioned on Day 1. Functional connectivity metrics were used to classify attention performance using various machine learning methods. Results revealed that combining the Adaboost model and recursive feature elimination yielded a classification accuracy of 91.84%. We discuss the implications of our results in developing neurofeedback frameworks to assess attention.

I. INTRODUCTION

Attention is a limited resource that allows us to focus on and employ cognitive capabilities to certain stimuli, enabling us to filter out extraneous information and prioritize what is essential [1]. Selective attention is the capacity to fixate on one stimulus while disregarding others. Divided attention, on the other hand, refers to the ability to devote attention to various things simultaneously, albeit at the expense of performance [1]. Technological advancements have shed light on the neurological foundations of attention mechanisms in recent years. Functional neuroimaging techniques such as functional magnetic resonance imaging (fMRI) and electroencephalography (EEG) have revealed information about the brain regions engaged in attentional activities. [1] The frontal cortex, particularly the anterior cingulate cortex (ACC) and the prefrontal cortex (PFC) is critical for attentional regulation, conflict monitoring, and goal-directed behavior orchestration. The parietal cortex, specifically the intraparietal sulcus (IPS), is involved in spatial attention orientation, allowing us to focus on specific regions in our visual field [1].

In the last few years, there has been a surge in interest in testing the efficacy of different methods to improve attention [2]. Non-invasive brain stimulation techniques like tDCS have become popular in the last decade. By delivering mild electrical currents across specific brain regions, tDCS modulates cortical excitability, eventually influencing neuronal activity. Investigations have revealed that tDCS applied over the dorsolateral prefrontal cortex (DLPFC), a crucial brain region modulating attentional mechanisms, increased attentional performance among healthy adult human participants [2,3]. Anodal stimulation of the DLPFC is believed to promote cortical excitability, resulting in an increased propensity of information processing and cognitive flexibility [3]. Anodal tDCS stimulation is expected to improve long-term potentiation (LTP), a learning and memory process. Nonetheless, cathodal tDCS stimulation has been demonstrated to increase LTD by affecting synaptic connections and brain networks that facilitate information processing and attentional mechanisms [2]. An empirical assessment of how long-term tDCS, i.e., tDCS administration over a period, impacts the neurophysiological basis of attention and the ensuing decision-making process is largely lacking and desperately needed in the literature.

Researchers in recent years have effectively amalgamated Electroencephalography (EEG) data with machine learning/deep learning techniques to derive significant insights on brain function and information processing and build predictive models for cognitive performance [5]. EEG is a non-invasive neuroimaging technique that uses electrodes on the scalp to analyze brain electrical activity. The most used techniques in EEG analysis include spectral power, event-related potentials (ERPs), connection measures, and various time-domain or frequency-domain metrics [5]. In recent years, there has been a lot of interest in source-space-based functional connectivity analysis for EEG [5,6]. The temporal correlation and synchronization of neuronal activity between brain regions is referred to as functional connectivity, and it provides insights into the interactions and networks underlying cognitive processes, behavior, and numerous neurological diseases [4].

Machine learning methods have been demonstrated to have the advantage of detecting granular associations and irregular patterns in EEG data that classic statistical methods might overlook [9]. Machine learning approaches such as support vector machines (SVM), gradient boosting machines (GBMs), and random forest (RF) have recently been used to identify lower-order cognitive processes such as workload, working memory, and so on, utilizing generic EEG parameters such as ERPs and spectral power [8]. For instance, researchers in [4] classified mental workload using phase-amplitude coupling and event-related spectral perturbation (ERSP), subsequently providing a paradigm for cross- and within-task workload discrimination. However, research on the efficacy of combining functional connectivity analyses and machine learning techniques to predict attention, particularly after tDCS administration, is lacking and desperately needed in the literature.

We intend to fill this research gap by utilizing EEG-based functional connectivity and machine learning algorithms to determine attention performance in a standard psychometric task following long-term tDCS administration. The following section summarizes previous research on predicting cognitive states and performance using EEG and machine learning techniques. The experiment and the functional connectivity metrics and machine learning models employed are explained subsequently. Finally, we discuss our findings and emphasize the necessity of precisely predicting changes in attention performance post-tDCS administration.

## II. BACKGROUND

EEG and machine learning algorithms have been utilized in recent years to predict changes in generic cognitive functions such as workload, attention, and working memory [17]. Machine learning approaches such as artificial neural networks (ANN), support vector machines (SVM), and random forest have been utilized to identify mental effort levels using EEG signals for binary (low- vs. high-workload levels) or multidimensional workload classification [5]. Despite this, most of the research has been focused on generic EEG features, resulting in low classification accuracies and lower generalizability and consistency across complex tasks [6-8]. Because cognitive processes vary considerably, proper classification across diverse cognitive activities, subjects, and sessions can be challenging. However, attention performance classification continues to be complex due to previous attempts yielding poor results. Researchers in [6], for example, employed ANN to achieve high binary classification accuracy of spatial attention performance, but subsequent multi-class classification generated substantially lower accuracy (less than chance) [6]. Researchers in [7] used regression to effectively forecast multi-class workload on two working memory (WM) tasks after their insufficient cross-task classification accuracy. However, assessing regression by mean square error is insufficient to compare with classification approaches [7] effectively.

Recent studies that use network modeling to depict the intricate interactions between brain areas provide an improved comprehension of the brain's functional structure by acquiring features, including dynamic reconstruction of functional connections between brain units [4]. The ability to uncover discriminatory network connections between distinct contexts in human cognition has been proved by increasing research merging classification algorithms with brain networks [9]. For instance, researchers in [10] explored the cross-frequency functional connectivity metric in a mental arithmetic task and found mental workload-related interactions between frontal theta and parieto-occipital higher alpha frequencies. Similarly, researchers in [9] used fMRI data to create CNN-based categorization based on functional connectivity to distinguish between distinct cognitive states. However, the literature has seldom explored forecasting attention performance using functional connectivity measures and machine learning algorithms, particularly following tDCS administration. In our research, we employed three functional connectivity metrics and different machine learning algorithms and feature selection methods to characterize and classify attention performance in a psychometric task following long-term tDCS administration.

## III. METHODS

### A. Experiment design

Forty participants (23 males, 17 females, mean age = 24.56 years, SD = 1.44 years) from the Indian Institute of Technology Mandi participated in the experiment. The Institutional ethical committee approved the study, and the experiment was conducted in strict compliance with the declaration of Helsinki. The experiment consisted of three phases: pre-intervention (Day 1), intervention (Day 2 to Day 8), and post-intervention (Day 10). All participants completed the pre-intervention phase on Day 1. They were informed about the experiment and its objectives and assessed for potential risks during tDCS administration using a screening questionnaire [2]. After obtaining written consent, demographic data was collected, and EEG data acquisition was conducted using a 32-channel system. The participants eventually executed the Macworth clock vigilance task (MCVT) for assessing attention performance with simultaneous EEG data recorded. Participants were then randomly divided into two conditions: experimental and placebo control. During simultaneous task training, the experimental group received 2mA anodal tDCS intervention for 15 minutes on the dorsolateral prefrontal cortex (dlPFC) every day from Day 2 to Day 8. A pre-and post-tDCS intervention questionnaire evaluated potential physical effects [2]. The control group received sham tDCS intervention (i.e., electrodes were placed but no tDCS was applied) during the same period. After a 24-hour break, participants executed the permuted rules operations task again (with simultaneous EEG being acquired) on Day 10. We utilized the Caputron Activadose 2 tDCS device to administer anodal tDCS, placing the anode on the left dorsolateral prefrontal cortex (dlPFC).

### B. Macworth clock vigilance test (MCVT)

The Macworth clock vigilance test is a psychometric task used to check attention and vigilance in various neurological and psychological studies [11]. The task was designed using Unity3D v2021.5.39. This task provides valuable insights into the dynamics of attention and the ability to sustain focus over time.

In this task, participants watch a red dot jump clockwise from one circular location to the next (24) at a steady speed. The red dot occasionally skips a place. Participants are prompted to hit the Spacebar button when they identify a missed event. The task takes place over a prolonged duration (30 minutes) to test the participant's sustained attention capacities. The task attempts to demonstrate the difficulties of sustaining focus over an extended time, mirroring real-world settings where humans must remain watchful and attentive to recognize infrequent but potentially crucial events. The red dot began at 12 o'clock on the computer interface and flashed in each "slot" in turn. Because the dot is visible for 0.65 seconds, each flash had a 0.65-second delay. With pseudorandom components, the signal events are "spread out" equally around the average event intervals.

We acquired several behavioral performance parameters in the Macworth clock vigilance test. Behavioral performance parameters acquired during the task included:

a) Number of hits (true positives) – Number of 'hits' (i.e., correct identification of a missed event) during the test

b) Number of false alarms (false positives) – Number of 'false alarms' (i.e., false identification of a perceived 'missed event' which is not a missed event) during the test

c) Number of false negatives (misses) – Number of missed events not identified by the participants

d) Number of true negatives (correct rejections) – Number of true events correctly rejected by the participants

e) Sensitivity (d') – Given as

$$d' = z\,(FA) - z(H) \ldots\ldots. (1)$$

where FA was the number of false alarms and H was the number of hits.

The sensitivity parameter was taken as the output variable to be predicted post-tDCS using different functional connectivity and machine learning algorithms. The difference in the percentage accuracy on Day 10 and Day 1 was calculated and used as the basis for eventual classification. The sensitivity parameter was found to be normally distributed; based on this condition, a three-class classification paradigm was designed based on the obtained mean (1.56) and standard deviation (0.11). The first class was labeled as low cognitive enhancement (range $\mu$ to $\mu+\sigma$ = 1.56 to 1.67), the second class as medium cognitive enhancement (range $\mu+\sigma$ to $\mu+2\sigma$ = 1.67 to 1.78), and the third class as high cognitive enhancement (range $\mu+2\sigma$ to $\mu+3\sigma$ = 1.78 to 1.89).

### C. EEG data acquisition and analysis

EEG data was obtained by employing 32 Ag/AgCl saline sensor electrodes inserted applying the standard 10-20 electrode placement method (Emotiv EPOC Flex and EMOTIVPRO data collection software, v2.34b, San Francisco, USA). The EEG data was captured at 256 Hz sampling rate. The electrode impedance was maintained below 10 K during the session. Anti-aliasing (0.1-45 Hz) with a band-pass filter and a 50 Hz notch filter was utilized to eliminate all the primary interferences. Before the commencement of the experiment, baseline EEG data (for 60 seconds) was acquired while the volunteers were requested to relax and keep their eyes open. Brainstorm [12] was used for EEG data pre-processing and feature extraction. It has been documented as a MATLAB plugin and is distributed without charge under the GNU General Public License [12]. Initially, the raw EEG data was band-pass filtered from 0.1 to 45 Hz. It was then re-referenced to the average of the electrodes in the left and right mastoid. Picard's Independent Component Analysis (ICA) was used to detect and remove eye blink artifacts [12]. The artifact-rejected EEG signal was subsequently modified with respect to the baseline data.

### D. Source localization

Using the symmetric boundary element method [13], we created a forward model for source localization of the 32 electrodes in the sensor space. The sensor-space, temporal EEG data was then projected into the forward model using dynamic statistical parametric mapping (dSPM) in the frequency domain. We used the Brainstorm plugin's standard ICBM 152 T1 image [13]. A volume conduction model was created using the default tissue conductivity parameters. The locations of electrodes on the scalp were mapped, and the sources were confined to grey matter. The power spectrum and cross-spectrum across EEG channels for the beta frequency band (13-29 Hz) were recovered using a Hanning window.

The current spectral density obtained was further parcellated into 28 different regions of interest (ROIs; 14 in each hemisphere) according to the Brodmann atlas [14]. The ROIs included the primary somatosensory area, primary motor areas (anterior/posterior), pre-motor area, Broca's area (pars opercularis and pars triangularis), primary/secondary visual area, the visual area in the middle temporal lobe, and the entorhinal and perirhinal cortex [14].

### E. Functional connectivity metrics

#### 1) Phase synchronization (PS)

The alignment of the phase angles of oscillatory impulses between two or more brain areas is called phase synchronization [14]. When the phases of these oscillations align or become "synchronized," it indicates that these regions are communicating with one another [14]. This coordination promotes information flow, allowing different brain areas to work together and increase the propensity to process information efficiently [14].

Funding agency: Life Sciences Research Board, Defence Research and Development Organization

Phase synchronization assumes that two oscillation mechanisms that do not have amplitude synchronization may possess phase synchronization. The Phase Locking Value (PLV) is the most widely used measure to determine phase synchronization strength [14]. Given a signal *a*, the instantaneous phase is given as:

$$\emptyset_x(t) = arctan \frac{\tilde{a}(t)}{a(t)} \dots \dots (2)$$

$\tilde{a}(t)$ is the Hilbert transform of *a(t)*, empirically defined as:

$$\tilde{a}(t) = \frac{1}{\pi} PV \int_{-\infty}^{+\infty} \frac{a(\tau)}{t-\tau} d\tau \dots \dots (3)$$

Where 'PV' is the Cauchy principal value. Therefore, the PLV between two ROIs is given as:

$$PLV = \left| \frac{1}{N} \sum_{j=0}^{N-1} e^{j(\emptyset_x(j\Delta t) - \emptyset_y(j\Delta t))} \right| \dots (4)$$

Where $\Delta t$ is the sampling period and *N* is the sample number of each ROI.

*2) Dynamic causal modeling (DCM)*

DCM goes beyond static metrics of connectivity to provide a more comprehensive understanding of the dynamic interplay between brain regions, shedding light on how they interact and impact one another over time [14]. DCM's theoretical roots assume that the brain operates as a complex network of interconnected regions, with one region influencing and being influenced by others [14]. It reveals the hidden causal architecture underlying reported patterns of cerebral activity by modeling brain-region interactions as a collection of differential equations that describe how activity in one region is influenced by activity in other regions [14].

DCM employs the concept of effective connectivity, defined as one neural system has influence over another. According to researchers in [14], the core concept of DCM is perceiving the brain as a deterministic non-linear dynamical system that receives inputs and creates outputs. A very accurate neural model of interconnected cortical areas is initially designed in DCM [14]. The forward model of how neural activity is translated into measurable responses is then added to this model. This allows the neural model's effective connectivity to be inferred from observed data.

*3) Wavelet coherence (WC)*

Wavelet coherence analyzes the temporal evolution of two signals over different frequency components, making identifying areas of strong synchronization or coupling easier [15]. In contrast to classic coherence approaches that rely on the Fourier transform, wavelet coherence provides a time-resolved and localized approach appropriate for capturing non-stationary and transient interactions [15].

The wavelet transform of a signal a is a function of both time and frequency [15]. It is given as the convolution of the input with a wavelet family $\varphi_u$:

$$W_a(t, f) = \int_{-\infty}^{\infty} a(u) \cdot \varphi_{t,f}^*(u) du \quad (3)$$

The cross-spectrum wavelet around time *t* and frequency *f* (given as input signals *a* and *b*) are derived from the wavelet transforms of *b* and *c*:

$$CSW_{AB}(t, f) = \int_{t-\frac{\delta}{2}}^{t+\frac{\delta}{2}} W_a(\tau, f) \cdot W_b^*(\tau, f) d\tau \quad (4)$$

where * represents the complex conjugate and $\delta$ is a frequency-dependent scalar. $\tau$ is the wavelet coefficient at time *t*. Therefore, the wavelet coherence at time *t* and frequency *f* is given as

$$WC_{AB}(t, f) = \frac{|CSW_{AB}(t, f)|}{|CSW_{AA}(t, f) \times CSW_{BB}(t, f)|^{\frac{1}{2}}} \quad (5)$$

We derived 28*28 adjacency matrices for each of the 40 participants (both for pre-and post-tDCS intervention) for the three functional connectivity metrics (PS, DCM, WC). These adjacency matrices were given as the input to the different machine learning algorithms. We also employed two different feature selection techniques to systematically explore the dataset (comprising of 784 features for each subject from each of the three functional connectivity metrics) for the most relevant and informative subset of features. Out of 784 features for each subject, the top 100 features were selected for further analysis for each functional connectivity metric. The feature selection techniques employed in this research work were:

*a) Pearson Correlation (PC):* Pearson correlation is a numerical value ranging from -1 to 1, where 0 represents no linear correlation, -1 represents perfect negative linear correlation, and 1 represents perfect positive linear correlation [16]. Regarding feature selection, the primary focus is on the relationship between each feature and the target variable [16]. A high positive correlation indicates that the target variable value grows as the feature value grows. A high negative correlation, on the other hand, indicates that when the feature value grows, the target variable value tends to drop. The top *n* features with the highest correlation coefficients are eventually selected [16].

*b) Recursive feature elimination (RFE):* The RFE method begins with training a model on all the features and assigns a weight or priority score to each feature depending on its contribution to the model's performance [16]. The features with the lowest weights or significance ratings is then pruned [16]. The model is retrained on the decreased feature set, and the feature reduction and model retraining procedure is repeated until a predetermined number of attributes or a stopping criterion is met [16].

*F. Machine learning models*

We employed different machine learning algorithms for classifying attention performance based on the functional connectivity algorithms explained in subsection E. The different machine learning algorithms used are as given below:

*a) Support Vector Machine (SVM):* SVM operates by determining the appropriate hyperplane for separating various classes in the data space [17]. The hyperplane is chosen to maximize the margin, which is the distance between the hyperplane and the nearest data points of each class, also known as support vectors [17].

*b) Decision Trees (DT):* A decision tree is built by recursively splitting data based on feature values to build subsets that are as pure as feasible, which means that each subset mostly comprises instances of a single class [17].

*c) Random Forest (RF):* The Random Forest approach generates many decision trees during training by selecting random subsets of the original dataset and random subsets of characteristics for each tree [17]. Each decision tree in the Random Forest is built using a technique known as recursive partitioning, which involves repeatedly splitting the data into subsets depending on the most discriminatory attributes, resulting in a tree-like structure [17].

*d) Multi-layer perceptron (MLP):* The MLP comprises numerous layers of linked neurons placed sequentially [17]. The layers are organized into three sections: an input layer that accepts input data, one or more hidden layers that do intermediary computations, and an output layer that produces final predictions or outputs [17].

*e) Adaboost:* AdaBoost (Adaptive Boosting) is a powerful ensemble learning technique that improves weak learners' predictive performance by integrating them into an effective, precise model. According to [17], the underlying principle behind AdaBoost is to train a sequence of weak learners, sometimes known as "base classifiers," consecutively and assign them weights based on their individual performance.

The different variations in the hyperparameters used in various machine learning algorithms are shown in Table 1. To pick the optimal parameters for the machine learning model, we utilized 10-fold cross-validation (three times) [17]. All the machine learning models were built in Python using scikit-learn. The machine learning models were trained for several hyperparameters, and the hyperparameters with the highest test accuracy during training folds were deemed the best. We employed grid search to determine each machine learning model's best set of hyperparameters. The machine learning models were trained for the various hyperparameters, and the hyperparameters with the highest accuracy were considered the best. The symmetric difference between the functional connectivity measures (PS, DCM, WC) on Day 1 and Day 10 was used as input to the machine learning models, and the difference in sensitivity parameters on Day 10 and Day 1 was taken as the output variable to be classified.

TABLE I. DIFFERENT HYPERPARAMETERS AND THE CORRESPONDING RANGE OF VALUES USED IN DIFFERENT MACHINE LEARNING ALGORITHMS

| Machine learning model | Hyperparameters varied |
|---|---|
| Support Vector Machine | 1) C – 0.01 to 100 (steps of 0.01)<br>2) $\gamma$ – 0.001 to 1 (steps of 0.001)<br>3) Kernel – Linear, polynomial, radial basis function |
| Decision Tree | 1) Maximum depth – 2 to 10 (steps of 1)<br>2) Minimum samples split – 2 to 10 (steps of 1)<br>3) Minimum samples leaf – 1 to 10 (steps of 1) |
| Random Forest | 1) Maximum depth – 2 to 10 (steps of 1)<br>2) Minimum samples split – 2 to 10 (steps of 1)<br>3) Minimum samples leaf – 1 to 10 (steps of 1)<br>4) Number of estimators – 10 to 100 (steps of 10) |
| Multi-layer perceptron | 1) Hidden layer sizes – 1 to 3 (steps of 1)<br>2) Hidden nodes count – 10 to 1000 (steps of 10)<br>3) Activation function – Logistic, tanh, rectified linear unit<br>4) Solver – Adam, stochastic gradient descent<br>5) Alpha - 0.0001 to 0.1 (steps of 0.001) |
| Adaboost | 1) Learning rate – 0.01 to 0.5 (steps of 0.01)<br>2) Maximum depth – 2 to 10 (steps of 1)<br>3) Subsample – 0.2 to 1.0 (steps of 0.2) |

IV. RESULTS

Table 2 shows the cross-validation test accuracies (in percentage) obtained for different functional connectivity algorithms combined with different feature selection techniques and machine learning algorithms.

TABLE II. TEST ACCURACY FOR THE DIFFERENCE IN THE SENSITIVITY PARAMETER BETWEEN DAY 1 AND DAY 10 OF THE MCV TASK FOR DIFFERENT FUNCTIONAL CONNECTIVITY METRICS, FEATURE SELECTION TECHNIQUES, AND MACHINE LEARNING ALGORITHMS

| Functional connectivity metric | Feature selection technique | SVM | DT | RF | MLP | Adaboost |
|---|---|---|---|---|---|---|
| **PS** | PC | 45.78% | 53.45% | 67.77% | 77.22% | 81.34% |
| | RFE | 52.56% | 58.78% | 69.81% | 78.66% | **91.84%** |
| **DCM** | PC | 57.89% | 56.75% | 64.45% | 76.33% | 83.28% |
| | RFE | 67.55% | 78.55% | 83.55% | 85.55% | 90.77% |
| **WC** | PC | 45.44% | 59.33% | 64.77% | 69.55% | 75.44% |
| | RFE | 58.55% | 63.44% | 69.55% | 72.23% | 84.56% |

Table 3 shows the best set of hyperparameters obtained during model calibration.

TABLE III. BEST SET OF HYPERPARAMETERS FOR DIFFERENT MACHINE LEARNING ALGORITHMS OBTAINED DURING MODEL CALIBRATION

| Machine learning model | Optimal hyperparameters during model calibration |
|---|---|
| SVM | - Kernel – linear, $\gamma$ = 0.001, C = 1 |
| DT | - Minimum samples split = 4, Minimum samples leaf = 8, Maximum depth = 2 |
| RF | - Number of estimators = 40, Minimum samples split = 5, Minimum samples leaf = 4, Maximum depth = 9 |
| MLP | - Solver – adam, hidden layer sizes = 50, $\alpha$ = 0.1, activation = reLu |
| Adaboost | - Learning rate = 0.1, subsample = 0.8, maximum depth = 3 |

As shown in Table 2, we obtained the highest three-class classification accuracy of 91.84% with the Adaboost model, combined with PS as the functional connectivity metric and RFE as the feature selection technique.

## V. DISCUSSION AND CONCLUSIONS

This research aimed to efficiently classify attention performance using functional connectivity metrics, feature selection techniques, and machine learning algorithms. Results revealed that PS-RFE-Adaboost model yielded a high classification accuracy of 91.84%. The significantly better classification results compared to other works [6-8] were consistent with [10], where they reasoned that due to the excellent spatial resolution and reduced volume conduction offered by dSPM (through the incorporation of boundary element methods and anatomical information), source estimation provided a more accurate and detailed representation of the underlying neural processes [10]. PS yielded a higher classification accuracy compared to other functional connectivity metrics. These results were consistent with [15], who elucidated that PS was able to extract significantly more information from the brain's functional networks. PS inherently provides directional information, indicating which brain regions causally influence the others. RFE and Adaboost well exploited these inherent advantages of using PS. Researchers in [16] had reasoned that RFE's propensity to handle multicollinearity and offer model agnosticism was preferred over other feature selection techniques for drowsiness detection [16]. This combined with Adaboost's capability to model complex and non-linear relationships, led to a higher classification accuracy in attention performance than the predecessors. However, this research work is not devoid of limitations. Even though the source localization technique employed a cortical parcellation technique (dSPM), it is still prone to volume conduction effects due to the inherent limitations in the EEG hardware. This disadvantage could be mitigated by employing an EEG acquisition system with more channels (>64) in the future. In addition, in the future, we intend to employ bleeding-edge deep neural networks that can interpret the spatial-temporal relationships between brain networks more efficiently. This might lead to a better understanding of the underlying cortical dynamics during attention tasks. This framework can potentially be used to design cognitive state assessors for real-time attention performance prediction post-tDCS administration.